# Highly stable linearly polarized arbitrary temporal shaping of picosecond laser pulses

## Fangming Liu[1,*], Senlin Huang[1], Kexin Liu[1], Shukui Zhang[2]

[1]Institute of Heavy Ion Physics, School of Physics, Peking University, Beijing 100871, China

[2]Thomas Jefferson National Accelerator Facility, 12000 Jefferson Avenue, Newport News, VA 23606, USA

### *Abstract*

This paper reports the study and demonstration of a new variable temporal shaping method capable of generating linearly polarized picosecond laser pulses with arbitrary predefined shapes, which are highly desired by various applications including low emittance high brightness electron bunch generation in photocathode guns. It is found that both high transmittance and high stability of the shaped pulses can be achieved simultaneously when birefringent stages (BSs) are set at specific phase delay. Such variable temporal shaping technique may lead to new opportunities for many potential applications over a wide range of laser wavelengths, pulse repetition rates, time structures and power levels, etc. In addition, a new double-pass variable temporal shaping method is also proposed and could significantly simplify the shaper structure and reduce the cost.

## Introduction

Laser temporal shaping plays an important role in many applications. For example, parabolic pulses are very useful for generation of super-continuum, ultra-short pulse, etc [1]. Flattop or 3D ellipsoidal pulses are much desired for reducing the emittance of electron bunches [2,3].

Due to short pulse duration and narrow spectral bandwidth, shaping picosecond pulse presents more difficulty than shaping pulses with duration in other time regimes [4]. Laser systems are often configured for a variety of laser wavelengths, pulse repetition rates, time structures, and power levels, high stability is often crucial for practical applications. Although many shaping methods are available in literature, versatile optical arbitrary waveform generation (VOAWG) applicable to the above concerns has been rarely reported thus far [4,5].

In this paper, a new variable laser temporal shaping method based on birefringent elements and polarizers is presented, featured with arbitrary shape generation and linear polarization, such a shaping technique shows good long term stability, and the capability of shaping pulses over a wide range of laser wavelengths with any pulse repetition rate and any time structure, and is very suitable for high power lasers.

## Variable shaper design

Figure 1 depicts the physical structure of the variable shaper to be reported in this paper. The layout is similar to a filter that has been used by astronomers in the study of the radiation spectrum from the sun [6,7]. Here, with the combination of coherent stacking concept [8] and precise optical phase control, variable shaping for arbitrary temporal profile is realized. In this temporal shaper, a number of identical birefringent stages (BSs) are placed between two polarizers labelled with #1 and #2, respectively. Although only eight BSs are depicted in Fig. 1, any number of BSs can be used. The slow axis and the fast axis of each birefringent stage (BS) are parallel with its end surfaces, which is also known as a-cut. In such a shaper, the rotation angles of all the BSs $(\Theta_1, \Theta_2, \cdots, \Theta_N)$ and output polarizer #2 $(\Theta_p)$ can be varied in full 360° for arbitrary pulse shaping, where $N$ is the total number of BSs.

Due to different group velocities between the laser pulses polarized along fast axis and slow axis of the BS, one linearly polarized laser pulse can split into $2^N$ mutually delayed replica pulses upon normally incidence on and passing through $N$ BSs and output polarizer #2 successively. The phase delay from each BS, which determines the phases of the replicas, is defined as $\varphi = 2\pi(cT/\lambda - \lfloor cT/\lambda \rfloor)$, where $\lfloor\, \rfloor$ is a floor function that omits decimal and only keeps integer (e.g., $\lfloor 10.8 \rfloor = 10$), $c$ is the speed of light in vacuum, $T$ is the time delay between two replicas produced by one BS, $\lambda$ is the laser wavelength. When all the BSs are identical and each introduces the same time delay $T$, the $2^N$ replicas can be grouped into $N+1$ groups to form $N+1$ replicas equally spaced in time, as shown in Fig. 2.

By rotating the BSs and the output polarizer #2 around laser propagation axis, the relative amplitudes of those $N+1$ replicas can be arbitrarily configured, therefore allowing output pulse with arbitrary profile to be produced. For any given output replicas' field amplitude vector $\boldsymbol{\mathcal{H}} = (H_1, H_2, \cdots, H_N, H_{N+1})$, at least one corresponding real $\boldsymbol{\Theta} = (\Theta_1, \Theta_2, \cdots, \Theta_N, \Theta_p)$ exists and can be calculated according to conservation of energy (by ignoring material absorption, the optical pulse energy remains the same at any point within the shaper [9]). $H_i$ $(i = 1,2, \cdots, N+1)$ can be any real number, and a negative value represents the vector element pointing to the opposite direction. After the rotation angles $\boldsymbol{\Theta}$ is determined, the replicas can be then calculated to produce the desired output pulse through coherent stacking.

*liu_fangming@pku.edu.cn

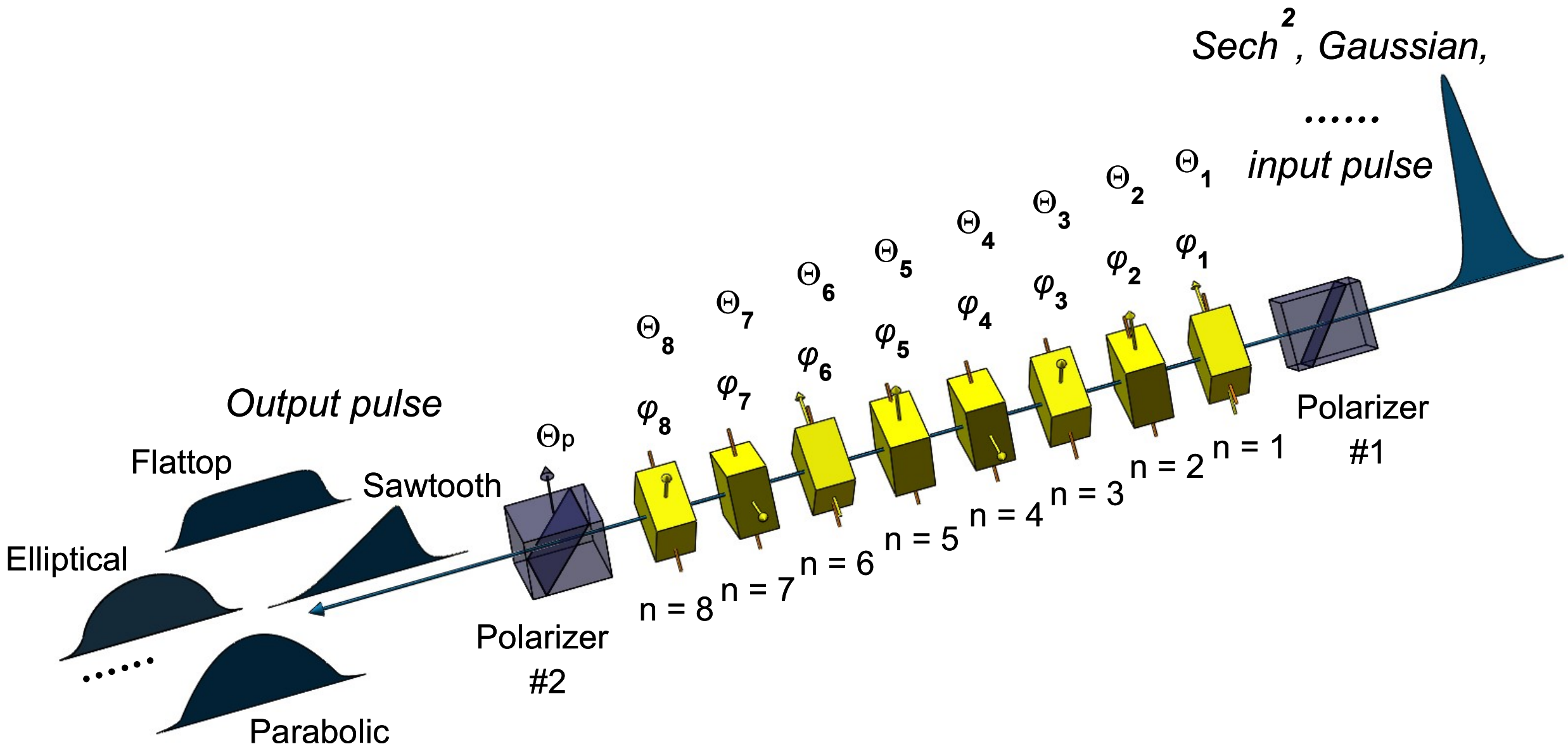

Fig. 1. Physical structure of the variable shaper for arbitrary temporal shaping. $\Theta_n$ is the angle between the slow axis of the *n-th* BS and the polarization direction of the input polarizer #1. $\Theta_p$ is the angle between polarization directions of the output polarizer #2 and the input polarizer #1. $\varphi_n$ is the phase delay of the *n-th* BS.

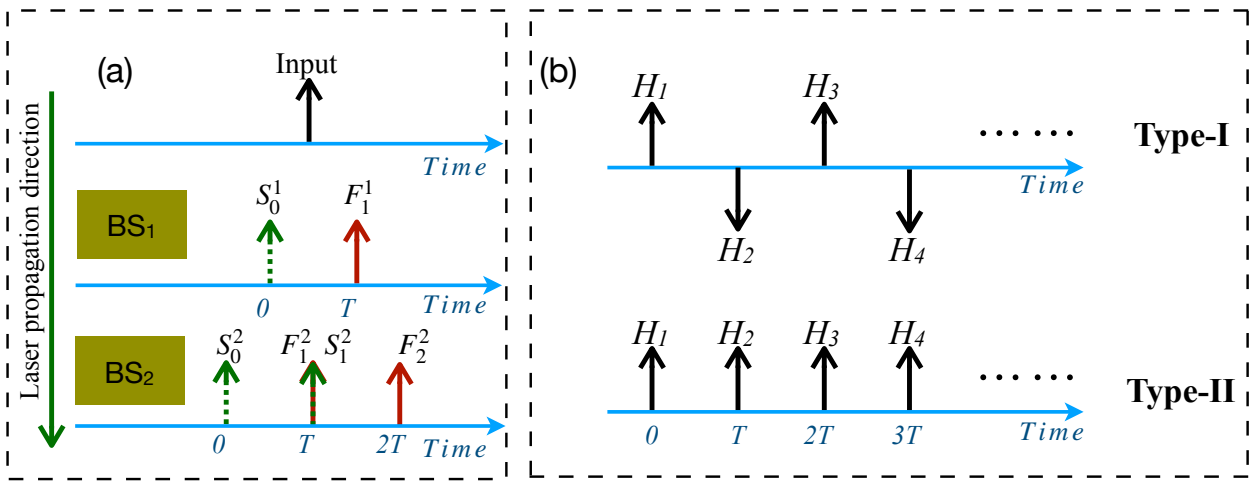

Fig. 2. (a) Replica pulses produced by one and two BSs ($BS_1$, $BS_2$). $S$ and $F$ denote the replicas that polarized along slow axis (green dashed arrows) and fast axis (red solid arrows) of the BS, respectively. The superscripts denote the BS sequence number, and the subscripts represent the replicas' time (in unit of $T$). (b) $N+1$ output replicas' field amplitude vector $\mathcal{H} = (H_1, H_2, \cdots, H_{N+1})$ with an equal interval $T$ between any adjacent components, where $|H_n|\ (n = 1,2,\cdots,N+1)$ represents the amplitude of the *n-th* replica field. In case of Type-I $\mathcal{H}$, $H_i * H_{i+1} < 0$ (i.e., all even-numbered vector elements point to one direction and all odd-numbered vector elements point to the opposite direction), while in case of Type-II $\mathcal{H}$, $H_i * H_{i+1} > 0$ (i.e., all $N+1$ vector elements point to the same direction), where $i = 1,2,\cdots,N$.

## Experiment and results

In the experiment, the input and output polarizer (#1, #2) are two Glan polarizers with extinction ratio on the order of $10^5$:1. Eight 3.3mm a-cut $YVO_4$ crystals were used as the BSs for shaping 532nm laser pulses with near *Sech*$^2$ temporal profile and a 6.5ps (FWHM) pulse width at 8.125MHz repetition rate. Each crystal was held inside a crystal oven where the temperature of each crystal can be independently controlled between 40°C ~ 180°C with an accuracy of 0.1°C.

One of the keys for realizing variable shaping with coherent stacking is to precisely control and stabilize the optical phase differences among replicas (i.e., the crystal phase delays in this case). It was found that the crystal phase delay is closely related to the transmittance as well as the stability of the shaping system [4,5]. Due to the dependence of the birefringence upon temperature and the thermal expansion of the crystal, the crystal phase delay can be fine-tuned between 0 ~ 2π by adjusting the temperature of the crystal. In the experiment, each crystal's phase delay-vs-temperature characteristics was individually calibrated by using two polarizers with their polarization directions parallel to each other [4].

All $YVO_4$ crystals and the output polarizer #2 were initially arranged in two ways (to satisfy Eqs. (1) or (2)), though such arrangements are not necessary for the arbitrary shaping system showed in Fig. 1.

$$\Theta_n = (-1)^n \frac{45^o}{N} + K * 90^o,\ \Theta_p = 90^o \quad (1)$$

$$\Theta_n = \frac{45^o}{N}(2n-1) + K * 90^o,\ \Theta_p = 0^o \quad (2)$$

Where $n = 1,2,\cdots,N$ and $K = 0,1$. Arranging shaper elements approximately to satisfy either Eqs. (1) or (2) is sufficient, but the main purpose is to choose a specific configuration: folded or a fan type [4,5,7]. There are two reasons for the way Eqs. (1) and (2) were set up. The first one is that such arrangements provide intuitional approaches for adjusting crystal rotation angles. Here assume $K = 1$ and the rules described below are similar when $K = 0$. In case of Eq. (1), when crystal sequence number $n$ is an odd (even) number, increasing $\Theta_n$ would decrease (increase) the amplitudes of the $n-th$ and the $(n+1)-th$ replica pulses at the same time increase (decrease) the amplitudes of all other $N-1$ replica pulses, and vice versa. While in case of Eq. (2), increasing $\Theta_n$ would increase the amplitude of the $n-th$ replica pulse and decrease the amplitude of the $(n+1)-th$ replica pulse, meanwhile exerting very small influence on the amplitudes of all other $N-1$ replica pulses, and vice versa. These rules are very useful in practice, and can be used as guides to directly tune crystal rotation angles for quickly achieving output pulse with a target profile [4,5].

The second reason is because they show that the shaper can be operated under two different modes with all BS phase delays set at $m * \pi$ to achieve high transmittance and high stability simultaneously through constructive interference, where $m = 1$ for vector Type-I $\boldsymbol{\mathcal{H}}$ (e.g., folded type shaper [4]), and $m = 0$ for vector Type-II $\boldsymbol{\mathcal{H}}$ (e.g., fan type shaper [5]). The shaper's output pulse varies versus the BS phase delay with a period of $2\pi$. The transmittance would drops to nearly zero due to destructive interference if all BS phase delays are close to $(m + 1) * \pi$.

With the help of real time cross-correlation measurement and guides from above described rules, several predefined target pulse shapes (e.g., parabolic, flattop, sawtooth, etc) were produced to demonstrate the arbitrary shaping capability of such variable shaper, the experimental examples are showed in Fig. X1 of the Appendix below, where very smooth profiles can be seen due to relative large ratio of input pulse width $T_1$ (6.5ps) versus crystal time delay $T$ (3.4ps), and all crystal phase delays being $m * \pi$. The Fig. X2 of the Appendix also presents more examples for demonstrating the arbitrary temporal shaping.

The stability of shaped pulse, which is very important for practical applications, also turned out to be excellent for such variable shaping method. Some experimental examples are showed in Fig. X3 of the Appendix. It is found that this high degree of stability is primarily due to the fact that crystal phase delays were fine tuned and maintained through precise control of crystal temperature, and were all set to $m * \pi$. Crystal phase delay at $m * \pi$ results in very small, nearly zero partial derivative of the output pulse intensity with respect to crystal temperature [4,5].

Besides generating pulses with arbitrary smooth profiles, such variable shaping technique can also produce various rippled pulse profiles and tunable pulse trains by adjusting the ratio $T_1/T$, as shown in Fig. 3 and Fig. X4 of the Appendix. The relative amplitudes of micropulses within the pulse train can be arbitrarily configured.

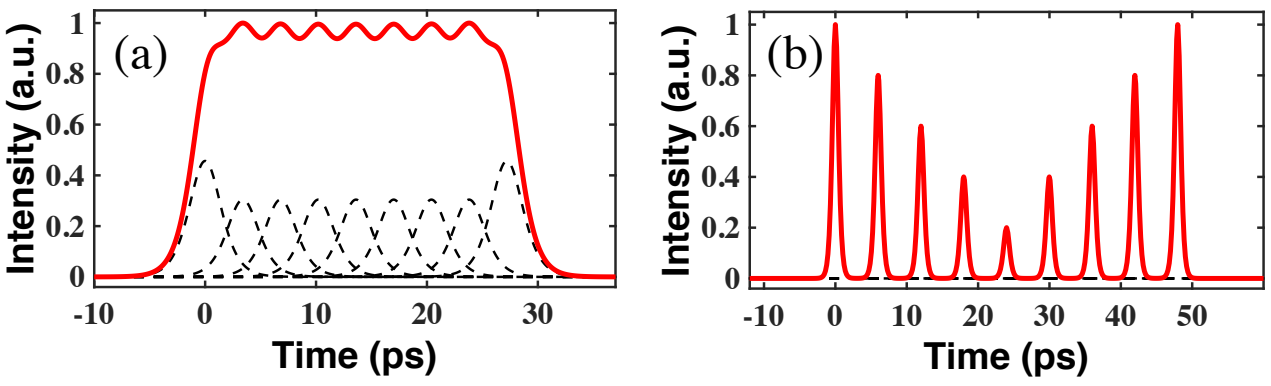

Fig. 3. Shaped output pulse (red solid line) and 8+1 replicas (black dashed lines). (a) Rippled pulse profile production with Type-I $|\boldsymbol{\mathcal{H}}|^2 = (1.5,1,1,1,1,1,1,1,1.5)$, $T_1/T = 1$, all BS phase delays are $\pi$. (b) Pulse train production with arbitrary predefined relative micropulse amplitudes (e.g., Type-II $|\boldsymbol{\mathcal{H}}|^2 = (5,4,3,2,1,2,3,4,5)$, $T_1/T = 1/6$, all BS phase delays are 0 rad in this case). Input pulse is 532nm laser with a *Sech²* temporal profile.

## Discussion

The configuration of the afore-mentioned variable temporal shaping technique for generating a specific pulse profile is not unique. For those $N + 1$ replicas that form the output pulse, vector $\boldsymbol{\mathcal{H}}$ of both Type-I (operating with all BS phase delays at $\pi$) and Type-II (operating with all BS phase delays at 0 rad) can produce the same pulse intensity profile as long as all $|H_i|$ $(i = 1,2,\cdots,N+1)$ are identical. Besides, for any given real vector $\boldsymbol{\mathcal{H}} = (H_1, H_2, \cdots, H_{N+1})$, it usually takes $2^{N/2+1} \sim 2^{N+1}$ of real $\boldsymbol{\Theta} = (\Theta_1, \Theta_2, \cdots, \Theta_N, \Theta_p)$ to realize it (e.g., there are 64 real $\boldsymbol{\Theta}$ for $\boldsymbol{\mathcal{H}} = (1,2,3,4,5,4,3,2,1)$). All those configurations are equivalent and lead to the same output pulse. If there are $\boldsymbol{\Theta}$ in which several adjacent BSs have the same or nearly same rotation angle value, these adjacent BSs may be replaced by one BS of the same total BS length. In addition, reversing the shaper would produce the same output pulse.

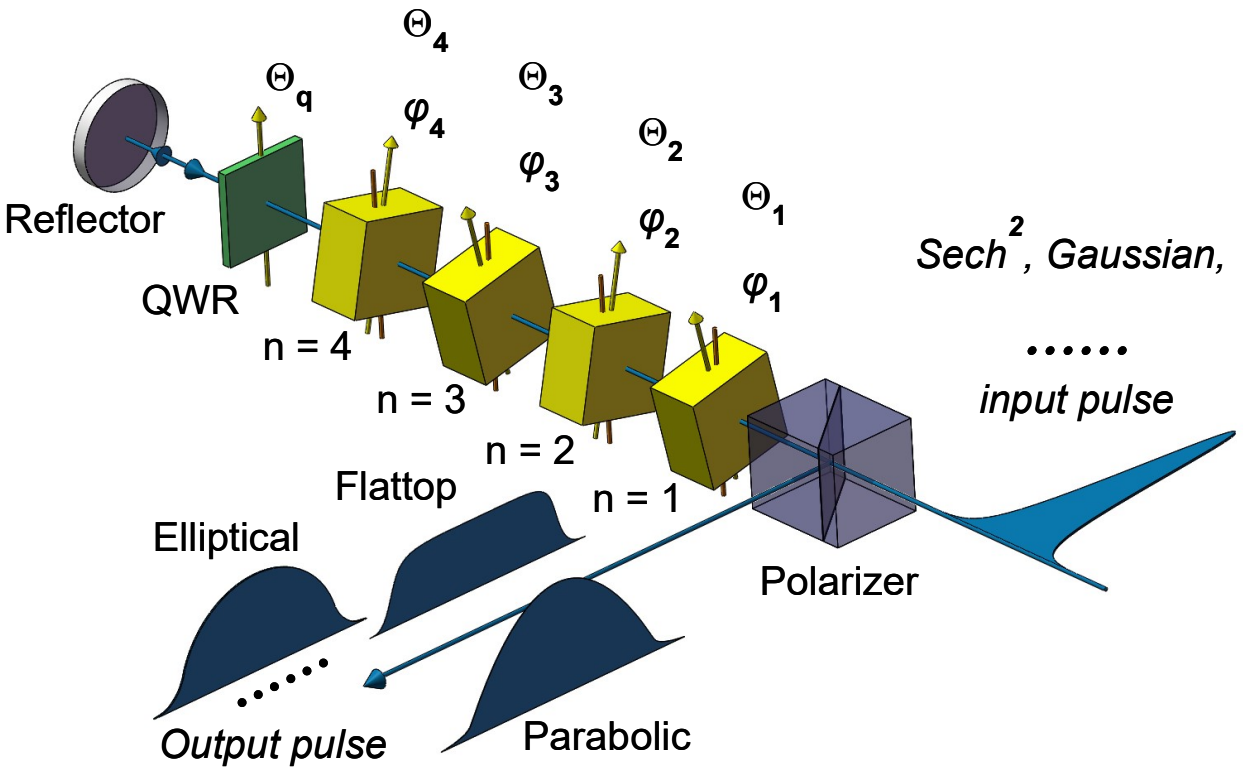

Fig. 4. Physical structure of the double-pass shaper (an example for vector Type-I $\boldsymbol{\mathcal{H}}$ is presented here, while Fig. X5 of the Appendix shows an example for double-pass shaper of vector Type-II $\boldsymbol{\mathcal{H}}$). $\Theta_q$, which denotes the angle between the fast or slow axis of QWR and the polarization direction of input polarizer, equals to $(\Theta_{N/2} + \Theta_{(N+2)/2})/2$ for a double-pass shaper, where $N$ is the total number of BSs required by single-pass shaper. The meaning of other symbols are the same with that in Fig. 1 above.

When $H_i = H_{N+2-i}$ $(i = 1,2,\cdots,N+1)$, the corresponding rotation angles $\boldsymbol{\Theta}$ would have $\Theta_{j+1} - \Theta_j = \Theta_{N+1-j} - \Theta_{N-j}$ $(j = 1,2,\cdots,N-1)$ and $\Theta_1 \pm 90^o = \Theta_p - \Theta_N$, and vice versa. At this point, for Type-I and Type-II output vector $\boldsymbol{\mathcal{H}}$ with $N$ being an even number, it is possible that the latter half of the single-pass shaper showed in Fig. 1 may be replaced by a quarter-wave-retarder (QWR) and a retro-reflector, resulting in a double-pass shaper for producing symmetrical pulses (e.g., parabolic, flattop, etc.), as shown in Fig. 4 and Fig. X5 of the Appendix. The shaped output pulse can be kicked out by the input polarizer. Since the laser pulse passes through each BS twice, the required number of BS by double-pass shaper can be reduced by half compared to that in a single-pass shaper [4,5].

Besides birefringent crystals, other birefringent materials such as electro-optic retarders [10] or fiber components [11,12] may also be used as BSs for the shaper, which would further extend the application range of such variable shaping technique. The maximum achievable transmittance is positive related to $1/N$ and the ratio $T_1/T$. More replica pulses can be easily realized by cascading more BSs, and would promote the shaping resolution (controlling the output pulse profile on a finer scale) but would not noticeably increase the complexity of the adjustment process and operation of the shaping system. The shaping systems introduced in [4,5] fall into the class of the more general shaping system depicted in Fig. 1.

For picosecond or sub-picosecond laser pulses, which have short pulse duration and narrow spectral bandwidth, such shaping method provide an efficient approach for arbitrary temporal shaping. In principle, such shaping method can be scaled approximately against the ratio $T/T_1$ for shaping lasers with pulse duration in femtosecond (fs) and nanosecond (ns) regimes, etc. Some examples are showed in Fig. X6 of the Appendix. For shaping fs pulse with broadband, it should be noted that the BS should be designed with careful consideration so that material dispersion and intensity related nonlinear effect won't severely distort the replica pulses. Since shaping for laser pulse with shorter duration needs shorter BS (smaller material dispersion), such characteristic may help alleviate the dispersion influences over ultrashort fs laser shaping with such kind of variable shaping technique.

## Summary

A variable temporal shaping method for generating arbitrary pulse shape with linear polarization is demonstrated. Arbitrary shape with smooth profiles, as well as rippled profiles and various tunable pulse trains can be produced by adjusting ratio $T_1/T$. It is discovered that high transmittance and high stability can be achieved simultaneously at specific BS phase delay. Such variable shaping technique shows good long term stability, and is robust, reproducible, easy for use and automation, capable of shaping laser pulses over a wide laser wavelength range (e.g., from IR to UV) with any pulse repetition rate and any time structure, and is directly applicable to high power lasers. A double-pass variable temporal shaping method is also proposed, featuring much more structure simplicity and cost reduction.

## Appendix

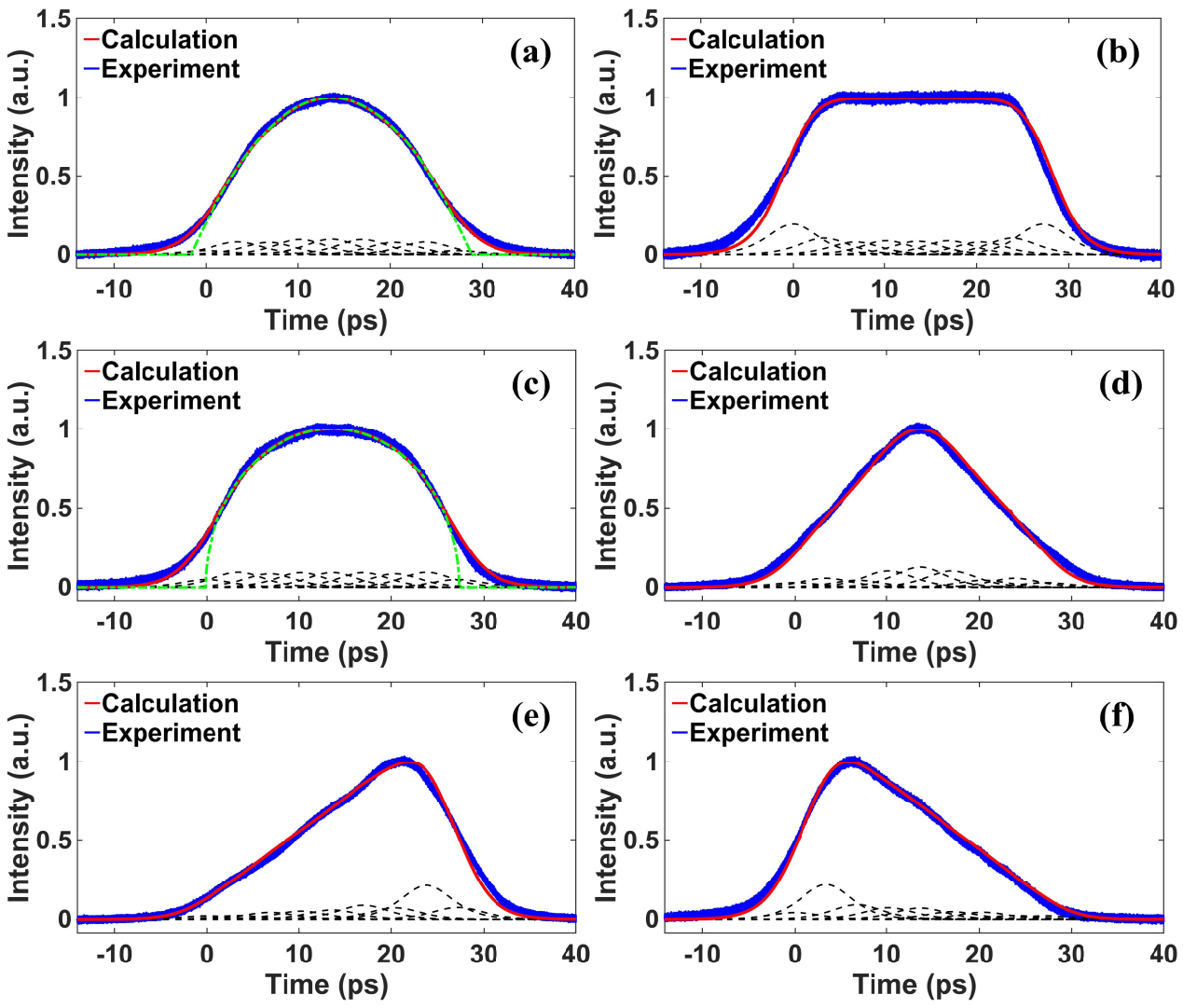

Fig. X1. Results of both the measured (blue lines) and calculated (red solid lines) pulse profiles after the shaper [4,5], (a) Parabolic, (b) Flattop, (c) Elliptical, (d) Triangular, (e) Sawtooth-I, (f) Sawtooth-II. Black dashed lines are theoretically calculated 8+1 replica pulses. Green dash-dotted lines are the ideal parabolic and elliptical curves. (a), (d) and (e) are vector Type-I $\mathcal{H}$ with all crystal phase delays set at $\pi$ rad, while (b), (c) and (f) are vector Type-II $\mathcal{H}$ with all crystal phase delays set at 0 rad. The laser beam transmittance, influenced by many factors, is measured to be ~20% in the experiment and is possible to achieve 40%~50% in theory with perfect BSs and polarizers for the shaper.

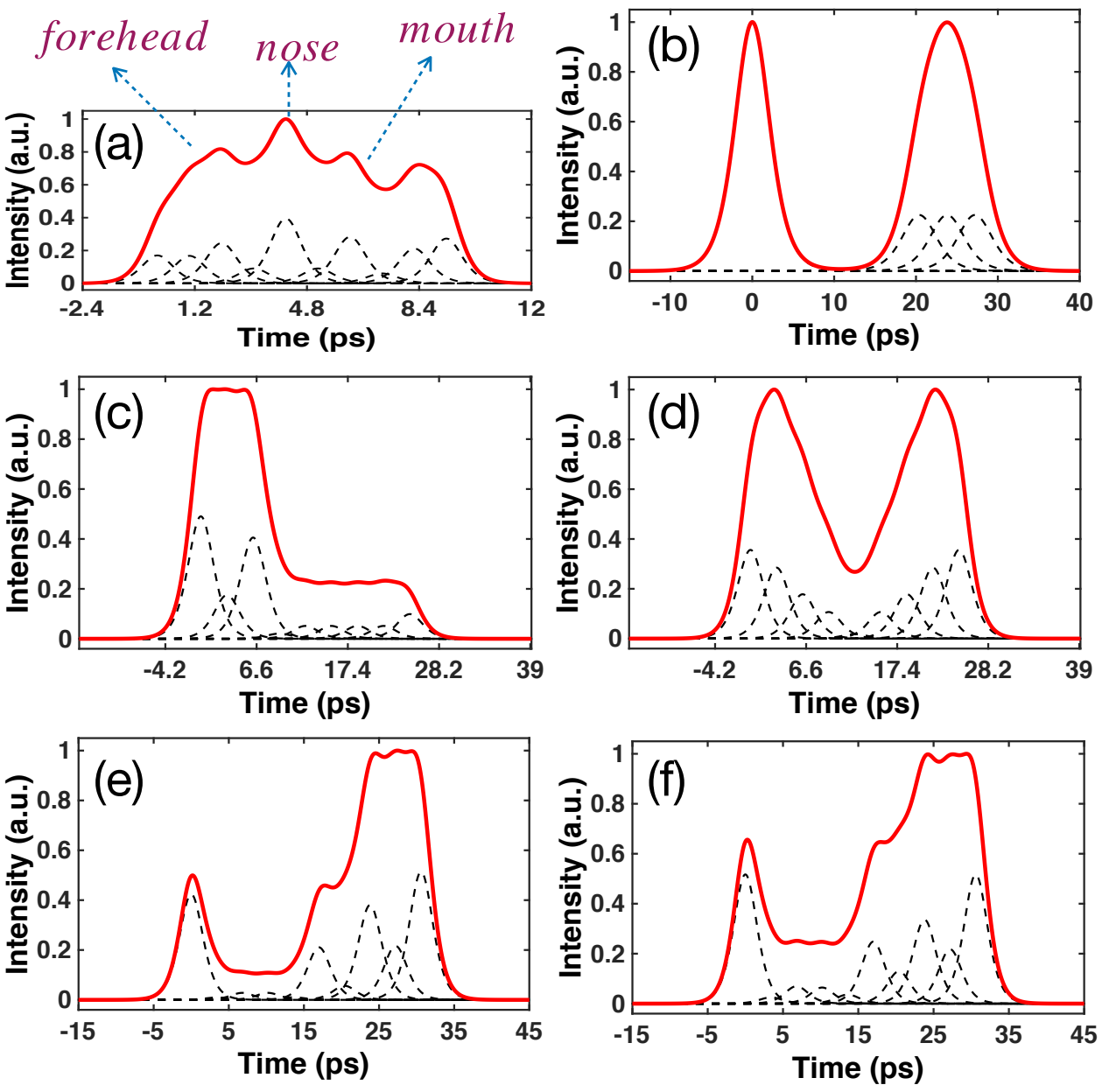

Fig. X2. Arbitrary laser pulse temporal profile production, e.g., (a) Human face profile, (b) Double pulses with different width, (c) English letter "L", (d) English letter "M", (e) Low-foot and (f) High-foot like profiles [13,14]. The meaning of all lines are the same with that in Fig. 3 above. The mirror profiles of those pulses can also be produced by this arbitrary temporal shaping method.

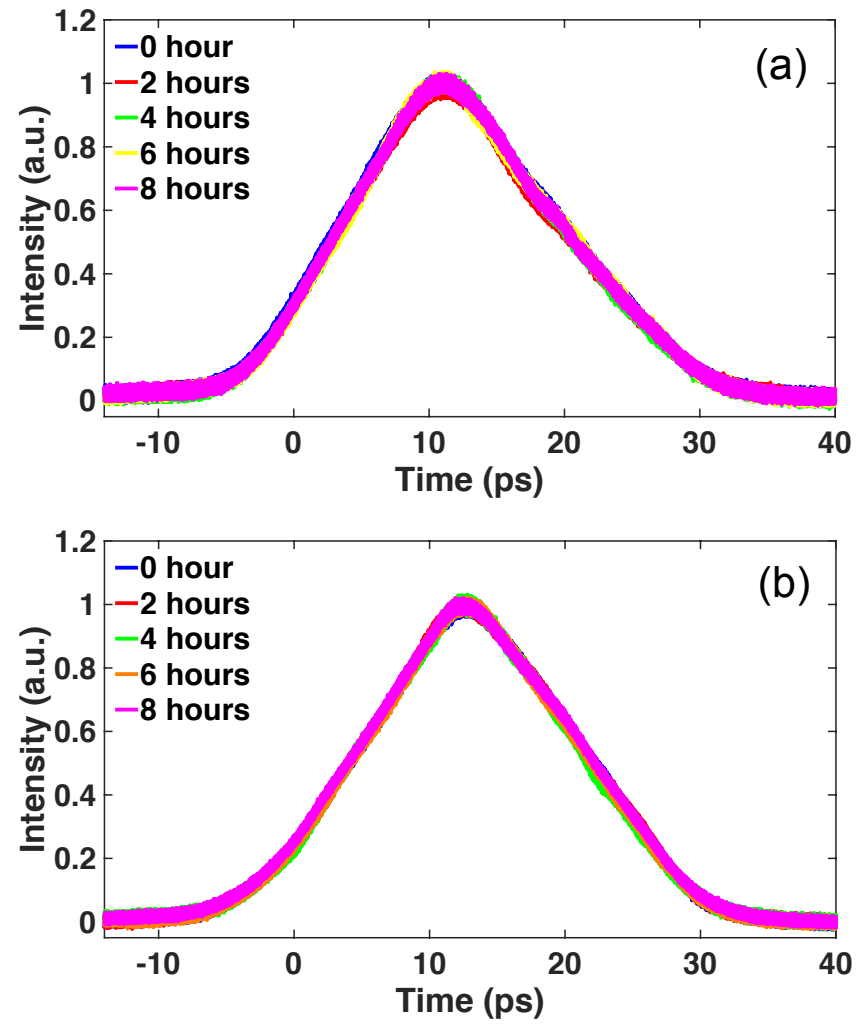

Fig. X3. Recording of measured triangular pulse shapes at different times during 8-hour continuous operation with all crystal phase delays set to (a) $\pi$ rad (vector Type-I $\mathcal{H}$) and (b) 0rad (vector Type-II $\mathcal{H}$) [4,5]. The laser beam transmittance through the shaper also remained unchanged, staying at about aforementioned 20% level.

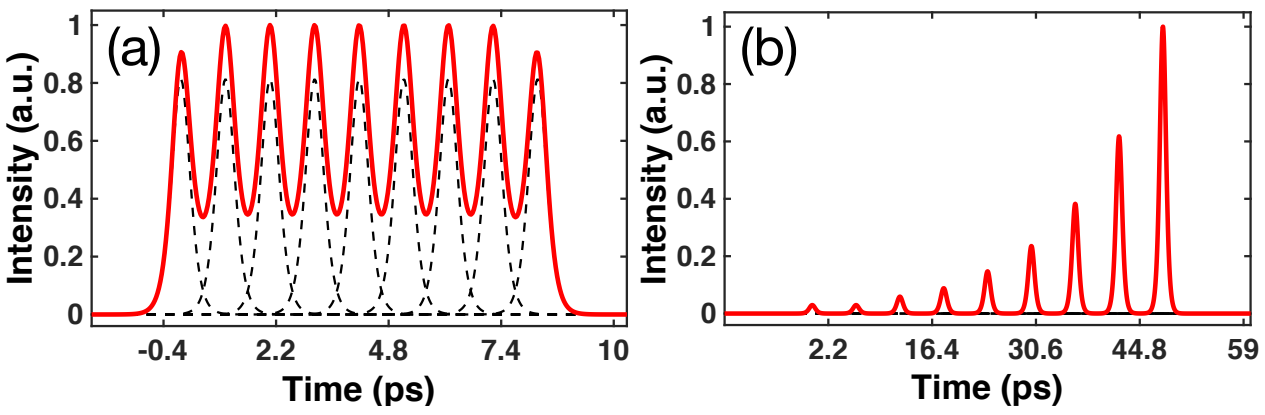

Fig. X4. (a) Rippled pulse production. The ripple depth can be adjusted by tuning the ratio $T_1/T$ (e.g., comparing to Fig. 3(a) showed above). (b) Pulse train production with arbitrary predefined relative micropulse amplitudes (e.g., Fibonacci sequence pulse train with $|\mathcal{H}|^2 =$ (1,1,2,3,5,8,13,21,34) in this case). The meaning of all lines are the same with that in Fig. 3 above.

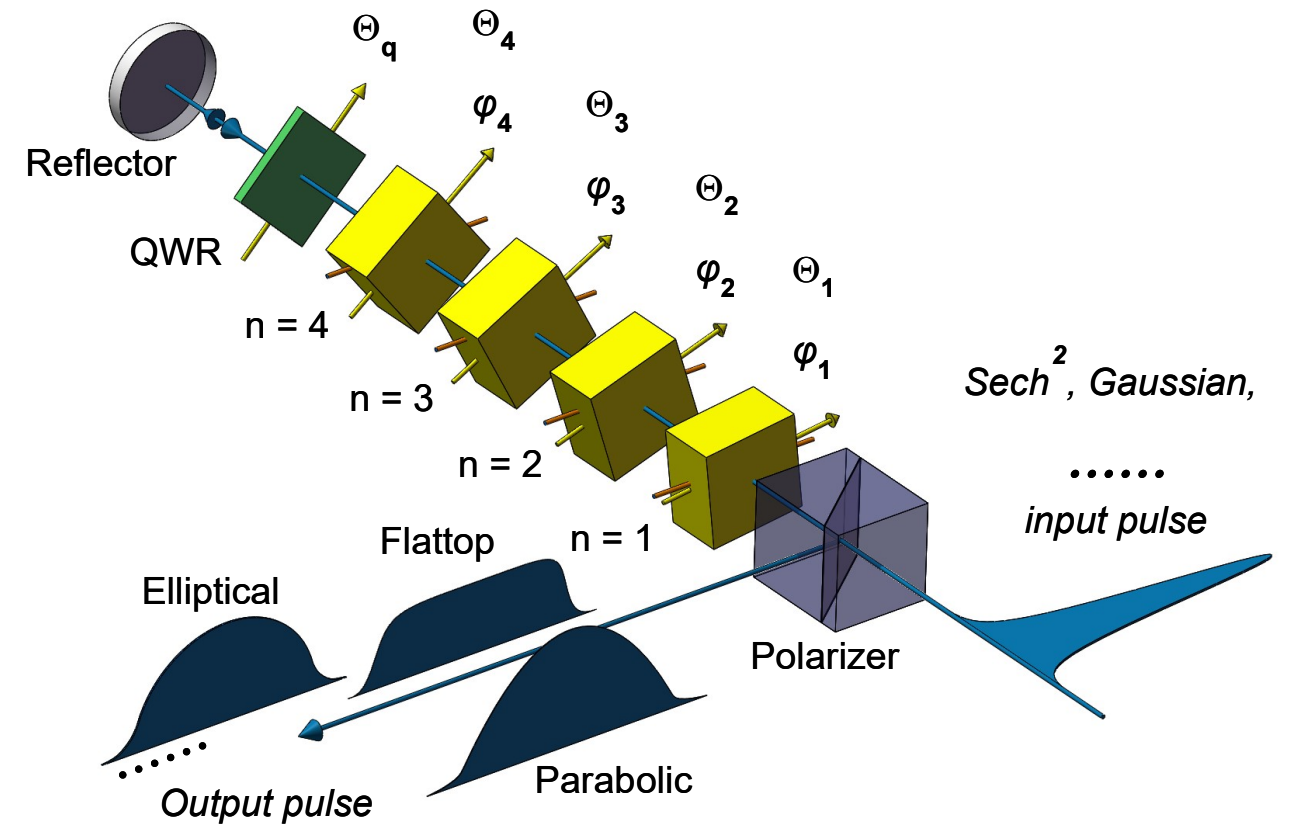

Fig. X5. Physical structure of the double-pass shaper (an example for vector Type-II $\mathcal{H}$ [5]). $\Theta_q$ equals to $(\Theta_{N/2} + \Theta_{(N+2)/2})/2$ for a double-pass shaper. The meaning of all symbols are the same with that in Fig. 4 above.

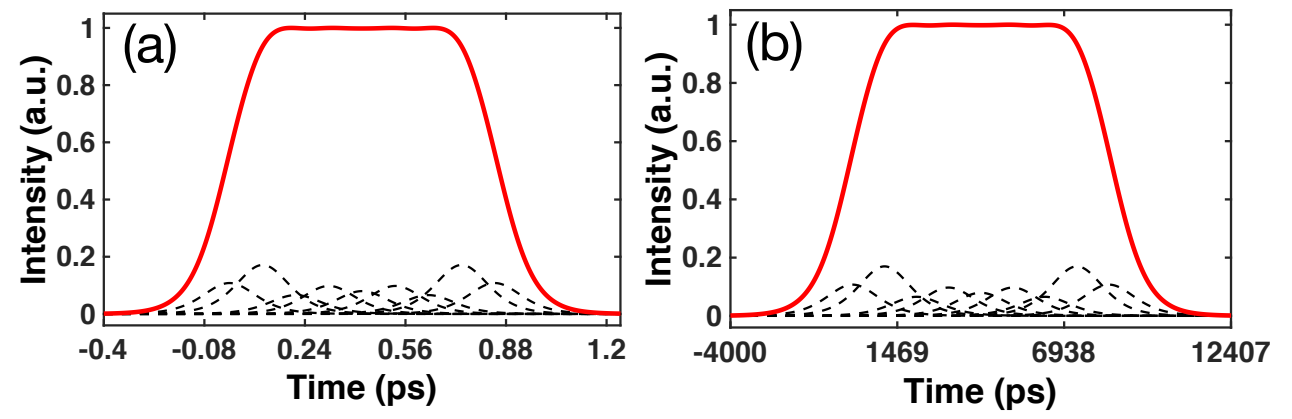

Fig. X6. Shaping laser pulse with initial duration in other time regimes, e.g., (a) femtosecond laser pulse, (b) nanosecond laser pulse, etc. The meaning of all lines are the same with that in Fig. 3 above.

## References

[1] F. Parmigiani, *et al*, "Ultra-flat SPM-broadened spectra in a highly nonlinear fiber using parabolic pulses formed in a fiber Bragg grating," Opt. Express **14**(17), 7617–7622 (2006).

[2] A. K. Sharma, T. Tsang, and T. Rao, "Theoretical and experimental study of passive spatiotemporal shaping of picosecond laser pulses," Phys. Rev. Spec. Top. Accel. Beams **12**(3), 033501 (2009).

[3] O. J. Luiten, S. B. van der Geer, M. J. de Loos, F. B. Kiewiet, and M. J. van der Wiel, "How to realize uniform three-dimensional ellipsoidal electron bunches," Phys. Rev. Lett. **93**(9), 094802 (2004).

[4] F. Liu, *et al*, "Generation of picosecond pulses with variable temporal profiles and linear polarization by coherent pulse stacking in a birefringent crystal shaper," Opt. Express **27**(2), 1467-1478 (2019).

[5] F. Liu, S. Huang, K. Liu, and S. Zhang, "Linearly polarized picosecond pulse shaping with variable profiles by a birefringent shaper." arXiv preprint arXiv:1901.06276 (2019).

[6] Z. Švestka, *et al*, "Solar Activity," Transactions of the International Astronomical Union, **14**(1), 71-110 (1970).

[7] I. Šolc, "Birefringent Chain Filters," J. Opt. Soc. Am. **55**(6), 621–625 (1965).

[8] S. Breitkopf, *et al*, "A concept for multiterawatt fibre lasers based on coherent pulse stacking in passive cavities," Light: Science & Applications **3**, e211 (2014).

[9] R. J. Pegis, "An Exact Design Method for Multilayer Dielectric Films," J. Opt. Soc. Am. **51**(11), 1255-1264 (1961).

[10] Y. Q. Lu, Z. L. Wan, Q. Wang, Y. X. Xi, N. B. Ming, "Electro-optic effect of periodically poled optical superlattice $LiNbO_3$ and its applications," Appl. Phys. Lett. **77**(23), 3719-3721 (2000).

[11] R. H. Chu, J. J. Zou, "Transverse strain sensing based on optical fibre Solc filter," Opt. Fiber. Technol. **16**(3) 151-155 (2010).

[12] Y. Yen, R. Ulrich, "Birefringent optical filters in single-mode fiber," Opt. Lett. **6**(6) 278-280 (1981).

[13] H. -S. Park, *et al*, "High-adiabat high-foot inertial confinement fusion implosion experiments on the national ignition facility," Phys. Rev. Lett. **112**(5), 055001 (2014).

[14] S. Atzeni, "Light for controlled fusion energy: A perspective on laser-driven inertial fusion," *EPL* **109** 45001 (2015).